*Article*

# Holographic Dark Information Energy

**Michael Paul Gough**

Space Science Centre, University of Sussex, Brighton, BN1 9QT, UK;
E-Mail: m.p.gough@sussex.ac.uk

**Abstract:** Landauer's principle and the Holographic principle are used to derive the holographic information energy contribution to the Universe. Information energy density has increased with star formation until sufficient to start accelerating the expansion of the universe. The resulting reduction in the rate of star formation due to the accelerated expansion may provide a feedback that limits the information energy density to a constant level. The characteristics of the universe's holographic information energy then closely match those required to explain dark energy and also answer the cosmic coincidence problem. Furthermore the era of acceleration will be clearly limited in time.

## 1. Introduction

Two of the fundamental principles of information theory are used here to argue that the dark energy causing the accelerating expansion of the Universe can be readily explained in terms of holographic information energy. The 'Holographic Principle' states that 'the complete set of degrees of freedom for all particles populating a certain region in space and time, can be represented as if they were all situated on the *boundary* of this space-time', 't Hooft [1]. There is one degree of freedom per elemental surface area, a square with a side equal to the Planck length, $l_p$, and then, crucially, the quantity of information in a region scales with the surface area bounding that region, rather than with the region's bulk volume [2,3]. Landauer's Principle states that any information erasure occurring in a system will necessarily cause that system to dissipate to its surroundings a specific minimum quantity of heat per erased bit. That quantity of heat depends solely on temperature, in accordance with the 2$^{nd}$ law of thermodynamics [4].

The Universe contains an amount of information that, via Landauer's Principle, has an equivalent energy and thus contributes to the total energy of the Universe. This information energy contribution is determined by the product of Universe information quantity in bits and the average energy equivalence of each bit. The former scales with the bounding area of the Universe from the holographic principle, while the later scales with average temperature via Landauer's Principle. Universe surface area and the average temperature of the dominant source of information energy are thus the key quantities needed to estimate the holographic information energy contribution to the Universe's energy balance.

## 2. Difference from Other Holographic Dark Energy Theories

There have been previous attempts to explain dark energy with theories that invoke the holographic principle, see, for example references [5–7]. As the holographic principle has its roots in string theory and quantum gravity theory these explanations are naturally highly theoretical in nature.

There is insufficient space here for a review of all these theories. Nevertheless, we can summarise previous holographic dark energy theories as providing a dark energy description in the form of a universe-wide vacuum energy, that, in some cases, is also equivalent to a cosmological constant. In the present work we estimate the energy equivalence of the holographic information of the Universe. Then the average information bit energy depends on the temperature of the strongest information energy contributions, shown below to be made by high temperature structure, stars, and stellar heated dust. Thus, although we assume overall information content scales with bounding area from the holographic principle, information energy in our model will not be evenly distributed within the 3-D bulk like a vacuum energy. Instead it is naturally associated with the location of high temperature structure. This clearly contrasts with the vacuum energy/cosmological constant of other holographic dark energy theories and is differentiated here by our choice of title: Holographic Dark Information Energy.

Dark energy accounts for the vast majority of the Universe's energy and it is this author's conviction that such a major phenomenon must be capable of being explained in terms of simple concepts. Accordingly, this article attempts to provide a straightforward explanation that quantitatively accounts for dark energy, and the resulting accelerating expansion of the Universe. We achieve this by taking a strictly phenomenological point of view, rather than by extending quantum gravity theory beyond the main statement of the holographic principle.

## 3. Holographic and Landauer Principles

It is well known that the surface area of a black hole determines its maximum entropy [8]. This discovery lead directly to the 'Holographic Principle'—all of the physical phenomena taking place within a 3-D region can be described on the surface bounding that region by one binary degree of freedom per unit Planck area ($l_p^2$) on that surface [2,3,9]. Although the holographic principle was originally invoked to describe particles in the vicinity of a Black Hole it is also considered to have universal validity. 'A full description of nature requires only a two-dimensional lattice at the spatial boundaries of the world', Susskind [3]. Basically the bulk 3-D description of a region can be fully translated into a boundary 2-D description even though the boundary has one less dimension. The two descriptions are completely equivalent. The holographic principle clearly derives its name from its similarity to the way a 2-D holographic image fully registers objects in 3-D. In this way our 3-D world is projected without information loss onto the 2-D screen [2,3]. It is thought information is conserved primarily on the 2-D outer surface of a region, possibly because that 2-D surface is the region's interface with the external world [10].

A black hole is the maximum entropy extreme with information content in bits equal to the number of Planck areas ($l_p^2$) needed to cover its surface area. To be precise, a black hole's entropy = ¼ area/ $l_p^2$ but the factor of ¼ is of little importance and ignored in our arguments here. The holographic principle is applicable to any object in the Universe, as well as to the Universe itself [3], so for regions, or systems, other than black holes their entropy, or information content, is well below this maximum level. For example, the total information content of the Universe is some 30 orders of magnitude less than the maximum entropy, or holographic bound, of the Universe. Nevertheless the holographic principle leads us to expect that the Universe information content will scale with Universe bounding area, as $a^2$, where $a$ is the scale size of the Universe, the size of the Universe normalized to the size today, $a = 1$.

Landauer's principle states that any erasure of information within a system must cause that system to dissipate $\Delta S\, T = k_B\, T\, ln2$ of heat per erased bit into the environment surrounding that system, increasing the entropy of that environment so that overall $\Delta I \geq 0$, [11–19]. This principle is based on the fact that information entropy and thermodynamic entropy are identical when the same degrees of freedom are considered. A bit of information is equivalent to $\Delta S = k_B\, ln2$ of thermodynamic entropy and the 2$^{nd}$ law of thermodynamics, $\Delta S \geq 0$, then states that total information, *I*, never decreases, $\Delta I \geq 0$. Although Landauer's principle is effectively only a restatement of the 2$^{nd}$ law, it is nevertheless a useful practical statement. Landauer's principle has allowed us to finally reconcile Maxwell's Demon with the 2$^{nd}$ law [20,21] and a recent experiment has directly verified this firm relation between information erasure and energy dissipation [22].

Landauer's principle applies equally to classical and quantum information and provides us with an energy equivalence for all information. This is similar to the standard cosmology practice of using $mc^2$ to represent the energy contribution of matter to the universe even though little mass has been converted to energy via nuclear fusion to date. Then, for any system, the information energy contribution is given by $N\, k_B\, T\, ln2$, where *T* the absolute temperature of that system, and *N* is the number of bits of information in that system in its 3-D bulk, or, completely equivalently, the number of bits in the holographic translation of that system on to its 2-D boundary.

## 4. Information Energy Contributions to the Universe

Table 1 lists the main contributions to the information of the Universe, amalgamating two reviews [23,24], and adding typical temperatures to estimate information energy contributions.

The low temperature relics of the Big Bang only provide very weak information energy contributions. Cosmic Microwave Background (CMB) has maintained effectively constant information since it decoupled from matter at the universe age of $3 \times 10^5$ years, with the CMB wavelength expanding in proportion to Universe size, cooling to the present temperature of 2.7 K. Relic neutrinos and gravitons decoupled at a very early time and have temperatures today ~2 K and ~0.6 K, respectively. We can also expect the dark matter contribution to be low if we assume a cold dark matter model.

**Table 1.** Information content, temperature, and information energy contributions.

|  |  | Information, N bits | Temperature, T °K | Information Energy $N k_B T \ln 2$, Joules |
|---|---|---|---|---|
| Relics of Big Bang | CMB photons | $10^{88}$–$2 \times 10^{89}$ | 2.7 | $3 \times 10^{65}$–$6 \times 10^{66}$ |
|  | Relic neutrinos | $10^{88}$–$5 \times 10^{89}$ | 2 | $2 \times 10^{65}$–$10^{67}$ |
|  | Relic gravitons | $10^{86}$–$6 \times 10^{87}$ | ~1? | $10^{63}$–$6 \times 10^{64}$ |
| Dark matter | Cold dark matter | ~$2 \times 10^{88}$ | $<10^{2}$? | $<10^{67}$ |
| Star formation | $10^{22}$ stars | $10^{79}$–$10^{81}$ | ~$10^{7}$ | $10^{63}$–$10^{65}$ |
|  | Stellar heated gas and dust | ~$10^{86}$ | ~$10^{6}$ | ~$10^{69}$ |
| Black Holes | Stellar sized BH | $10^{97}$–$6 \times 10^{97}$ | ~$10^{-7}$ | $10^{67}$–$6 \times 10^{67}$ |
|  | Super massive BH | $10^{102}$–$3 \times 10^{104}$ | ~$10^{-14}$ | $10^{65}$–$3 \times 10^{67}$ |
| Universe | Holographic bound | ~$10^{124}$ | - | - |

Although there is considerable uncertainty in these figures, we note that stellar heated gas and dust has an information energy two orders of magnitude greater than the relics of the Big Bang and dark matter, and also at least an order of magnitude greater than black holes. In the remaining sections we concentrate solely on this strongest contribution and it is therefore necessary to justify our omission of black hole effects before continuing.

Single super massive black holes (~$10^7$ solar masses) that occur at the centre of most galaxies and other stellar sized black holes dominate the information bit content contributions to the universe (left-hand data column of Table 1). Classically a black hole has a temperature at absolute zero emitting no radiation, but, in quantum theory, Hawking radiation is emitted with a perfect Planck spectrum that corresponds to an extremely cold object. For example, a stellar sized black hole typically has a temperature of only ~$10^{-7}$ of a degree above absolute zero. The larger super massive black holes at galactic centres have even lower temperatures ~$10^{-14}$ degrees. Table 1 shows that, despite their massive information contents, black holes do not over dominate the information energy contributions (right-hand column of Table 1) as much on account of their ultra-low temperatures. Furthermore, this black hole information remains inside the black hole's event horizon and thus invisible to, and unavailable to, the Universe as a whole. The so called 'no hair theorem' states that, from the Universe's point of view, a black hole can be *fully* specified by just three parameters: mass; angular momentum; and charge. Any two black holes with the same mass, angular momentum, and charge will be *absolutely* indistinguishable from each other and thus their apparent information content, as seen by the universe, must be negligible.

Solar mass sized black holes have lifetimes ~$10^{64}$ years, very many orders of magnitude greater than the present age of the Universe. It is most likely that they eventually evaporate away via Hawking radiation. In this way their information energy slowly leaks out but, at any one time, at least at the present time, stellar sized black holes exhibit negligible information content to the rest of the Universe. How, and whether, all of the massive internal information content ever returns to the Universe at the end of a black hole's lifetime, the 'black hole information paradox', is therefore not relevant to our discussion here. Super massive black holes at galactic centres have even longer lifetimes but those in active galactic nuclei do provide the Universe with some information by way of sporadic energetic outbursts of high radiation fluxes emanating from inwards falling matter. But a recent survey of $10^5$ galaxies has shown that these black holes spend most of their time dormant with only 1.6% active at any one time [25]. At the small size extreme, the only black holes with lifetimes as short as the present Universe age might be primordial black holes of mass $\sim 10^{15}$ grams formed immediately after the big bang. These should evaporate emitting photons of ~100 MeV before finally exploding with higher energy TeV photons [26]. However, despite being proposed as one possible source of gamma-ray bursts, there is little observational evidence to date to support even the existence of such primordial black holes [26]. So, overall, we can conclude that the Universe today is generally unaware of the information that resides effectively locked away in black holes of *all* possible sizes.

Finally, it is very difficult to take gravitational entropy into account here in any separate quantitative way because so little is known about it [27]. As structure and stars form in the Universe the ensuing reduction in conventional entropy is believed to be accompanied by a much larger increase of gravitational entropy that ensures *ΔS ≥ 0* [27]. Increasing gravitational entropy is also accompanied by an increase in information which will, in turn, also have an equivalent energy per bit determined solely by temperature according to Landauer's principle. In the following sections our use of the average universe baryon temperature for stellar heated gas and dust therefore provides a representative temperature for both the thermodynamic and gravitational information associated with structure and star formation. Thus gravitational entropy is effectively included in the work below.

## 5. Information Energy from Stellar Heated Gas and Dust

Table 1 and the above arguments show that stellar heated gas and dust at a present mean gas temperature of $\sim 2 \times 10^6$ K [28] provides the main information energy contribution today. We can expect this contribution to have varied over time with the rate of star formation and to have been proportional to the fraction of all baryons that are in stars. Figure 1 collects together many sources [28–37] of measurement and various models for the average baryon temperature, *T*, and the fraction of all baryons that are found in stars, *f*, and plots this data as a function of universe scale size, *a*. Universe scale size, *a*, is related to redshift, *z*, by $a = 1/(1 + z)$. While we usually think of the universe primarily as cooling over time with the low temperature relics of the big bang, the average temperature of baryons is seen in Figure 1 to have increased significantly with increasing integrated star formation in recent times.

All of the data and models in Figure 1. can be characterized with sufficient accuracy for our purposes by simple power laws with a distinct change from a dependency that goes as $a^{+3}$ at times before $z \sim 1$ ($a < 0.5$) to one that goes as $a^{+1}$ afterwards. As the average baryon temperature varies as $a^{+1}$ in this recent period the average energy equivalence per bit will also have varied as $a^{+1}$ from Landauer's principle. At the same time we expect the total number of bits in the Universe to have increased with Universe bounding area as $a^{+2}$ from the holographic principle. Thus the total holographic information energy will have increased as $a^{+3}$ while the volume also increased as $a^{+3}$ to give an overall constant information energy density during this recent period $z < 1$, ($a > 0.5$).

**Figure 1.** Summary of models and measurements for the average baryon temperature, *T*, and the fraction of all baryons in stars, *f*, as a function of Universe scale size, *a*, (relative to now, *a* = 1). We assume that all of the above measurements and models can be represented with sufficient accuracy for our purposes by the *grey lines*, corresponding to the specific functions $a^{+1}$ and $a^{+3}$ for the time periods $z < 1$ and $z > 1$ respectively. Sources of data plotted- **Combined models** (*continuous black line*): mean gas temperature using GADGET model, Nagamine *et al.* [28], modelled stellar densities, Hopkins & Beacom, [29]; and Le Borgne *et al.*, [30]. **Individual measurements** (symbols): *open rectangles* Dickinson et al, [31]; *open stars* Brinchmann & Ellis, [32]; *filled circle and rectangles* Rudnick *et al.* [33]; *triangles* Cohen [34]; *Crosses* Perez-Gonzalez *et al.* [35] and Elsner *et al.*, [36]; *open circle* Galactic fraction today, Cole *et al.*, [37].

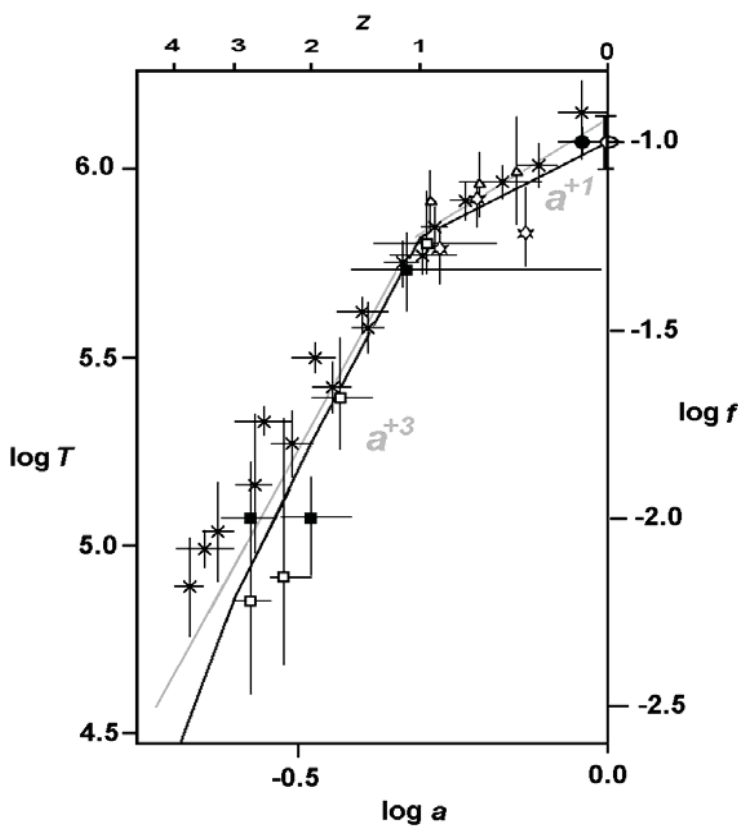

All other Universe components have energy densities that naturally decrease as the Universe expands. These components are usually characterised by their equation of state parameter, *w*, a useful comparator against the matter energy density and defined by the component's energy density variation as $a^{-3(1+w)}$. While the equation of state parameters, $w_m = 0$ for matter varying as $a^{-3}$, and $w_r = +1/3$ for electromagnetic radiation varying as $a^{-4}$, only the apparently constant dark energy ($a^0$) has been found to exert a negative pressure with $w_{de} \sim -1$. In fact any component with a negative equation of state parameter must contribute to dark energy. From the above, holographic information appears to naturally have a constant energy density, or an equation of state parameter very close to $w_i = -1$, during the period $a > 0.5$, which, coincidentally, is also the period when dark energy has acted on the universe.

The average Universe baryon temperature is presently ~2 × 10⁶ k (~10% of typical stellar temperatures since $f \sim 0.1$) and corresponds to an average bit energy around 120 eV. Dark energy in the standard flat cosmological model needs an energy density today 3 × the mass energy density to explain the acceleration in Universe expansion. Then, for holographic information energy to account for all dark energy, the total information content of the Universe would need to be ~$10^{87}$ bits, a value one decade above the survey of Table 1 but below previous estimates ~$10^{90}$ bits [38–40]. This similarity in value and the fact that holographic information has an equation of state parameter, $w_i = -1$, at the crucial time of an accelerating Universe, makes holographic information a very strong contender to explain dark energy.

## 6. Resulting Accelerating Expansion

The previous section showed that holographic information could easily account for the present dark energy density value, even though there is still some uncertainty in the estimate of Universe information content. Accordingly, in the following sections we start with the assumption that the holographic information energy density does indeed provide all of the dark energy, with a value today three times the present mass energy density, corresponding to the standard constant flat cosmological model. Then we can look at the time history to see if it is also consistent with the observed effects of dark energy.

In particular, the change in baryon temperature gradient at $a = 0.5$ in Figure 1 from scaling as $a^{+3}$ before to $a^{+1}$ afterwards may lead to some significant difference in behaviour. Prior to $a = 0.5$ the holographic total information energy scaled as $a^{+5}$, since the information content again scaled as $a^{+2}$, and thus the holographic information energy density scaled $a^{+2}$, for an equation of state parameter, $w_i = -1.67$. So, while holographic information energy has the same equation of state as dark energy, $w_i = -1.0$, during recent times, $a > 0.5$, it clearly has not acted over all time like a cosmological constant, one of the possible explanations for dark energy. We need to see whether the lower information energy density at earlier times will lead to a detectably different time profile for the accelerating expansion.

First we plot in Figure 2, upper panel, the matter, holographic, and total energy densities over time, assuming a present holographic energy density value three times that of the matter energy density, corresponding to the standard constant flat cosmological model. We note that, because of the much higher mass energy density at earlier times, there is very little difference between the total energy (continuous black line), sum of mass (blue line) plus holographic information energy(continuous red line), and the total energy (dashed black line), sum of mass plus a cosmological constant of the same value (dashed red line).

At present the only way to detect dark energy is via the resulting accelerating expansion. So we wish to ascertain whether there will be any significant difference in this resulting accelerating expansion between the action of holographic information energy and the action of a true cosmological constant. For the present dark energy density value three times the matter energy density, the Hubble parameter, $H$, is a factor of two greater today than it would have been if the universe had expanded without dark energy. This can be seen from $3H^2 \approx 4 \times 8\pi G\rho$ with dark energy instead of $3H^2 \approx 8\pi G\rho$ without, where $G$ is the Gravitational constant and $\rho$ the Universe mass density, including dark matter [41]. In Figure 2, lower panel, we plot the expected accelerated expansions from holographic information (continuous red line) and a cosmological constant (red dashed line). As we are primarily concerned with the difference between them, these have been simply estimated from the square root of the ratio of total energy in each case to matter energy. Clearly the two red lines produce very similar accelerations. For comparison, and to justify our simple approach, we also plot (grey line) the standard flat cosmological model for a dark energy 3 × the mass energy, from the inverse of the growth rate of large scale cosmic structure [42].

>**Figure 2.** Plotted against the logarithm of the universe scale size (size now, *a* = 1) are: Upper panel: Logarithm of energy densities normalised to the present total energy density(= 1): blue line, mass energy; red line, holographic information energy; black line, sum of mass and holographic; dashed black line, sum of mass and a cosmological constant (red dashed line). Lower panel: Relative extra expansion, *a\*/a*, caused by accelerating expansion: red line, accelerating expansion from holographic energy; red dashed line, accelerating expansion for a cosmological constant; grey line, the standard constant flat cosmological model (dark energy = 3 × mass energy); and, green line, estimated minimum accelerating expansion required to provide the observed change in star formation rate dependency from $a^3$ to $a^1$ at $z \sim 1$ in Figure 1.

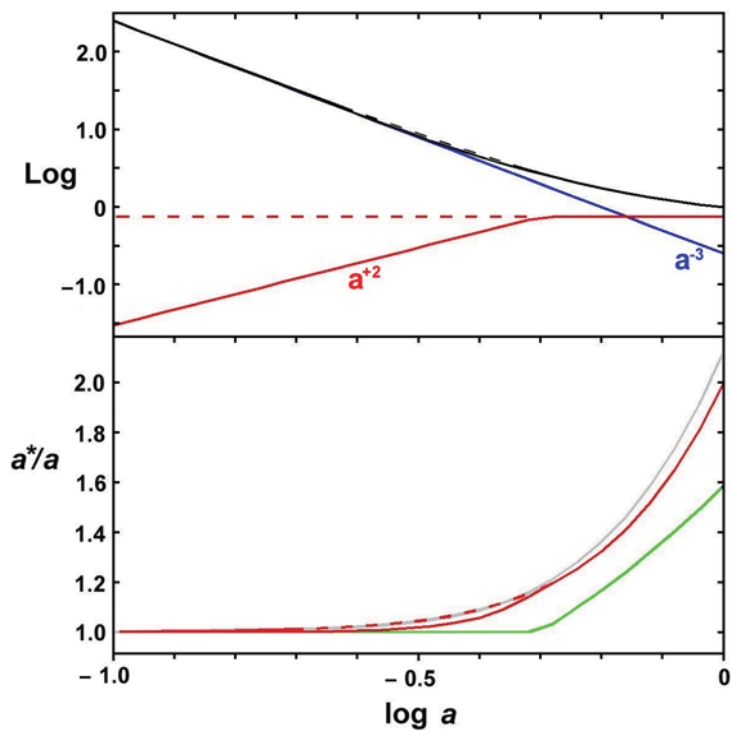

We conclude that there is little significant difference between the effects of holographic information energy and a cosmological constant. In future, when the accelerating expansion has been measured more accurately we might be able to detect the slightly lower holographic acceleration at earlier times in the region of log *a* ~ −0.5. Also the change in star formation gradients (Figure 1) provides a weak cusp or bulge at *a* ~ 0.5 (log *a* ~ −0.3) that might be detectable, but we note that this feature might have been artificially accentuated by our simple representation of the data by just two power laws.

Overall, it will be very difficult to distinguish by observation between holographic energy and a cosmological constant as being the cause of the accelerating expansion. We note that a survey of the latest acceleration measurements is most consistent with the type of holographic dark energy model that replicates a cosmological constant [43]. These recent measurements are therefore also generally consistent with our Holographic Dark Information Energy model because of the very small difference in resultant acceleration between our model and a cosmological constant (see Figure 2).

## 7. Possible Cosmic Feedback

Now we must ask why does the star formation rate conveniently change at z ~ 1 in Figure 1 to the specific rate that provides a constant holographic energy density? Previously this change in the rate of star formation, also manifested by an associated reduction in the rate of assembly of large scale structure in the Universe at this time, has been directly attributed to the accelerating expansion [43]. We do not wish to go into the complex physics of star formation here but we can assume that the more rapidly decreasing Universe matter density caused by the acceleration will lead to an overall reduction in star formation rate. We can make a rough estimate of the minimum extra expansion needed to produce the observed change in gradient. In Figure 2 lower panel, green line, we plot the specific scaling of $a$ that will give the ratio of matter densities equal to the ratio between the earlier $a^{+3}$ dependency in Figure 1 extrapolated to $z < 1$ and the actual $a^{+1}$ values at that time. This will be the minimum required expansion as there will be a tendency for smaller scale structure to continue to form stars more or less independent of the extra acceleration of the wider Universe. This minimum extra expansion (green line) has a strong similarity in form and value to the holographic (red line) and standard model (grey line).

Thus the change in star formation gradient in Figure 1 at $a = 0.5$ appears to result naturally from holographic information starting to provide an energy component large enough to cause the accelerating expansion. At $a = 0.5$ the mass density, scaling as $a^{-3}$ would have been eight times the present mass density while the holographic energy density had just levelled out at today's value, three times the present mass energy density. Then holographic information was just starting to make itself felt with an energy density ~3/8 of the mass energy density at the transition in gradient. The earlier, faster $a^{+3}$ rate of temperature increase, corresponding to information energy density increasing as $a^{+2}$, could not continue beyond $a = 0.5$ because it would have lead to an even faster accelerated expansion. This would have given a more rapidly reducing matter density that, in turn, would have caused the star formation rate, and thus the holographic energy density, to fall. Thus we suggest that there is a clear feedback from the acceleration that reduces the rate of star formation and hence limits acceleration driving force, the holographic information energy, to a constant level.

There appears to be a natural balance between acceleration and star formation. The present $a^{+1}$ temperature variation providing a constant dark information energy density is therefore a naturally preferred balancing state that will continue as long as the fraction of all nucleons in stars, $f$, can continue to increase at this rate. Presently $f \sim 0.1$, but with $f$ clearly limited by definition to below unity, we can expect this balance and therefore the acceleration in expansion eventually to cease, certainly before $a = 10$.

Figure 2 used the specific information content that gives a present holographic information energy density three times the mass energy density. Note, however, that the steepness of information energy density rise prior to the change in gradient (as $a^{+2}$) and the steepness of mass energy density fall (as $a^{-3}$) leads to an overall behaviour that has a relatively low dependence on the precise information content value. For example, if the universe information content had been as much as an order of magnitude higher, or lower, then the equivalent time at which information energy density reached 3/8 of mass energy density would have occurred at $a \sim 0.3$, or $a \sim 0.8$, respectively. In either scenario we would still be living today in an era dominated by the holographic dark information energy that, if the above feedback hypothesis is correct, would still be maintained at a constant energy density.

Finally, we argued above that black hole information is effectively locked away in ultra-low temperature, opaque-walled vaults and does not contribute significantly to the Universe information energy. Nevertheless black holes have played a significant role. Table 1 shows that if that large quantity of information had instead been made available to the Universe, even at just the few degrees Universe background temperature, then the acceleration would have started so much earlier that the sun and many of the stars we see today could never have formed!

## 8. Cosmic Coincidence and Summary

We have shown that our approach of combining the holographic principle with Landauer's principle accounts for the present constant dark energy density, explains the timing of dark energy's emergence as a significant force in recent cosmic history, and also obviates the need for a true cosmological constant.

Holographic dark information energy firmly ties the accelerating Universe expansion to the recent history of star formation, to answer the cosmic coincidence question—'Why now?' The present constant dark energy density is probably a natural balance between accelerating Universe expansion and the rates of large scale structure assembly and star formation.

Arguments presented here, based directly on the Holographic principle and Landauer's principle, depending solely on Universe bounding area and average baryon temperature, may seem overly simple. However, they are supported by a consistent phenomenological fit to observations, and also, of course, by Occam's razor. Finally, the results of this approach provide strong support for the universal applicability of the holographic principle.

## References

1. 't Hooft, G. Obstacles on the way towards the quantization of space, time and matter- and possible solutions. *Stud. Hist. Phil. Mod. Phys.* **2001**, *32*, 157–180.
2. 't Hooft, G. Dimensional reduction in quantum gravity. In *Salamfestschrift: a Collection of Talks*; *World Scientific Series in 20th Century Physics*; World Scientific Publishing Co.: Hackensack, NJ, USA, 1993; Volume 4, e-print gr-qc/9310026.
3. Susskind, L. The world as a hologram, *J. Math. Phys.* **1995**, *36*, 6377.
4. Landauer, R. Irreversibility and heat generation in the computing process. *IBM J. Res. Dev.* **1961**, *3*, 183–191.
5. Li, M. A model of holographic dark energy. *Phys. Lett. B*, **2004**, *603*, 1–5.

6. Gong, Y.; Wang, B.; Zhang, Y-Z. Holographic dark energy reexamined. *Phys. Rev. D* **2005**, *72*, 043510:1–043510:6.
7. Shalyt-Margolin, A. Entropy in the present and early universe: new small parameters and the dark energy problem. *Entropy* **2010**, *12*, 932–952.
8. Bekenstein, J.D. Black holes and entropy. *Phys. Rev. D* **1973**, *7*, 2333–2346.
9. Bousso, R. The holographic principle. *Rev. Mod. Phys*. **2002**, *74*, 825–874.
10. 't Hooft, G. The fundamental nature of space and time. In *Approaches to Quantum Gravity, Toward a New Understanding of Space, Time and Matter*; Oriti, D., Ed.; Cambridge Univ. Press: Cambridge, UK, 2009; pp. 13–25.
11. Bennett, C.H. Logical reversibility of computation. *IBM J. Res. Dev.* **1973**, *17*, 525–532.
12. Bennett, C.H. The thermodynamics of computation—A review. *Int. J. Theor. Phys*. **1982**, *21*, 905–940.
13. Bennett, C.H. Notes on the history of reversible computation. *IBM J. Res. Dev*. **1988**, *32*, 16–23.
14. Landauer, R. Dissipation and noise immunity in computation and communication. *Nature* **1988**, *335*, 779–784.
15. Landauer, R. Computation: A fundamental physical view. *Phys. Scr*. **1987**, *35*, 88–95.
16. Feynman, R.P. *Lectures on Computation*; Penguin Books: London, UK, 1999; pp. 137–184.
17. Ladyman, J.; Presnell, S.; Short, A.J.; Groisman, B. The connection between logical and thermodynamic irreversibility. *Stud. Hist. Phil. Mod. Phys*. **2007**, *38*, 58–79.
18. Piechocinska, B. Information erasure. *Phys. Rev. A* **2000**, *61*, 062314:1–062314:9.
19. Daffertshofer, A.; Plastino, A.R. Landauer's principle and the conservation of information. *Phys. Lett. A* **2005**, *342*, 213–216.
20. Bennett, C.H. Notes on Landauer's principle, reversible computation, and Maxwell's demon. *Stud. Hist. Phil. Mod. Phys*. **2003**, *34*, 501–510.
21. Leff, H.S.; Rex, A.F.; Eds. *Maxwell's Demon 2: Entropy, Classical and Quantum Information, Computing*. IOP Publishing Ltd: London, UK, 2003
22. Toyabe, S.; Sagawa, T.; Ueda, M.; Muneyuki, E.; Sano, M. Experimental demonstration of information-to-energy conversion and validation of the generalized Jarzynski equality. *Nat. Physics* **2010**, *6*, 988–992.
23. Frampton, P.H.; Hsu, S.D.H; Reeb, D.; Kephart, T.W. What is the entropy of the universe? *arXiv* **2008**, arXiv:0801.1847v3.
24. Egan, C.A.; Lineweaver, C.H. A larger estimate of the entropy of the universe. *arXiv* **2009**, arXiv:0909.3983v1.
25. Haggard, D.; Green, P.J.; Anderson, S.F.; Constantin, A.; Aldcroft, T.L.; Kim, D-W.; Barkhouse, W.A. The field Xray AGN fraction to z = 0.7 from the CHANDRA multiwavelength project and the Sloan Digital Sky. *Astrophys. J*. **2010**, *723*, 1447–1472.
26. Carr, B.J. Primordial black holes as a probe of cosmology and high energy physics. *arXiv* **2003**, arXiv:astro-ph/0310838v1.
27. Penrose, R. *The Road to Reality*; Jonathan Cape: London, UK, 2004; pp. 705–707.
28. Nagamine, K.; Loeb, A. Future evolution of the intergalactic medium in a universe dominated by a cosmological constant. *New Astron.* **2004**, *9*, 573–583.

29. Hopkins, A.M.; Beacom, J.K. On the normalisation of the cosmic star formation history. *Astrophys. J.* **2006**, *651*, 142–154.
30. Le Borgne, D.; Elbaz, D.; Ocvirk, P.; Pichon, C. Cosmic star formation history from a non-parametric inversion of infrared galaxy counts, *Astron. Astrophys.* **2009**, *504*, 727–740.
31. Dickinson, M.; Papovich, C.; Ferguson, H.C.; Budavari, T. The evolution of the global stellar mass density at 0 < z < 3. *Astrophys. J.* **2003**, *587*, 25–40.
32. Brinchmann, J.; Ellis, R.S. The mass assembly and star formation characteristics of field galaxies of known morphology. *Astrophys. J.* **2000**, *536*, L77–L80.
33. Rudnick, G.; Rix, H-W.; Franx, M.; Labbe, I.; Blanton, M.; Daddi, E.; Foerster Schreiber, N.M.; Moorland, A.; Rottgering, H.; Trujillo, I.; van de Wel, A.; van der Werf, P.; van Dokkum, P.G.; van Starkenburg, L. The rest-frame optical luminosity density, color, and stellar mass density of the universe from z = 0 to z = 3. *Astrophys. J.* **2003**, *599*, 847–864.
34. Cohen, J.G. CALTECH faint galaxy redshift survey. XVI, The luminosity function for galaxies in the region of the Hubble Deep Field-North to z = 1.5. *Astrophys. J.* **2002**, *567*, 672–701.
35. Perez-Gonzalez, P.G.; Rieke, G.H.; Villar, V.; Barro, G.; Blaylock, M.; Egami, E.; Gallego, J.; Gil de Paz, A.; Pascual, S.; Zamorano, J.; Donley, J.L. The Stellar Mass Assembly of Galaxies from z = 0 to z = 4: Analysis of a Sample Selected in the Rest-Frame Near-Infrared with Spitzer. *Astrophys. J.* **2008**, *675*, 234.
36. Elsner, F.; Feulner, G.; Hopp, U. The impact of Spitzer infrared data on stellar mass estimates—and a revised galaxy stellar mass function at 0 < z < 5. *Astrophys. J.* **2008**, *477*, 503.
37. Cole, S.; Norberg, P.; Baugh, C.M.; Frenk, C.S.; Bland-Hawthorn, J.; Bridges, T.; Cannon, R.; Colless, M.; Collins, C.; Couch, W.; *et al*. The 2dF galaxy redshift survey:near-infrared galaxy luminosity functions. *Mon. Not. R. Astron. Soc.* **2001**, *326*, 255–273.
38. Lloyd, S. Ultimate physical limits to computation. *Nature* **2000**, *406*, 1047–1054.
39. Lloyd, S. *Programming the Universe: A Quantum Computer Scientist Takes On the Cosmos*; Alfred A. Knopf Publisher: New York, NY, USA, 2006.
40. Lloyd, S. Computational capacity of the universe. *Phys. Rev. Lett*. **2002**, *88*, 237901:1–237901:4.
41. Räsänen, S. The effect of structure formation on the expansion of the universe, *Int. J. Mod. Phys* **2008**, *D17*, 2543–2548.
42. Guzzo, L.; Pierleoni, M.; Meneux, B.; Branchini, E.; Le Fèvre, O.; Marinoni, C.; Garilli, B.; Blaizot, J.; De Lucia, G.; Pollo, A.; *et al*. A test of the nature of cosmic acceleration using galaxy redshift distortions. *Nature* **2008**, *451*, 541–544.
43. Li, M.; Li, X.; Zhang, X. Comparison of dark energy models: A perspective from the latest observational data. *Sci. China G* **2010**, *53*, 1631–1645.